\documentclass[%
 amsmath,amssymb,
 reprint,%
prb,
superscriptaddress,
longbibliography,
]{revtex4-1}

\usepackage{graphicx}
\usepackage{dcolumn}
\usepackage{bm}
\usepackage{hyperref}
\usepackage{xcolor}
\usepackage{soul,color}
\usepackage{verbatim}
\hypersetup{
    colorlinks=true,
    linkcolor=blue,
    citecolor=blue,
    urlcolor=blue,
}
\setcitestyle{numbers,square}

\begin{document}

\preprint{APS/123-QED}

\title[manuscript in preparation]{Mutual synchronization of constriction-based spin Hall nano-oscillators in weak~in-plane~fields}

\author{Hamid~Mazraati}
\affiliation{NanOsc AB, Kista 164 40, Sweden}
\affiliation{ 
Department of Applied Physics, School of Engineering Sciences, KTH Royal Institute of Technology, Electrum 229, SE-16440 Kista, Sweden
}%
\author{Shreyas~Muralidhar}
\affiliation{Department of Physics, University of Gothenburg, 412 96, Gothenburg, Sweden}

\author{Seyyed~Ruhollah~Etesami}
\affiliation{Department of Physics, University of Gothenburg, 412 96, Gothenburg, Sweden}

\author{Mohammad~Zahedinejad}
\affiliation{Department of Physics, University of Gothenburg, 412 96, Gothenburg, Sweden}

\author{Seyed~Amirhossein~Banuazizi}

\affiliation{ 
Department of Applied Physics, School of Engineering Sciences, KTH Royal Institute of Technology, Electrum 229, SE-16440 Kista, Sweden}%

\author{Sunjae~Chung}
\affiliation{ 
Department of Applied Physics, School of Engineering Sciences, KTH Royal Institute of Technology, Electrum 229, SE-16440 Kista, Sweden}%
\affiliation
{Department of Physics, University of Gothenburg, 412 96, Gothenburg, Sweden}

\author{Ahmad~A.~Awad}
\affiliation{NanOsc AB, Kista 164 40, Sweden}
\affiliation{Department of Physics, University of Gothenburg, 412 96, Gothenburg, Sweden}

\author{Mykola~Dvornik}
\affiliation{NanOsc AB, Kista 164 40, Sweden}
\affiliation{Department of Physics, University of Gothenburg, 412 96, Gothenburg, Sweden}

\author{Johan~\AA{}kerman}
\affiliation{NanOsc AB, Kista 164 40, Sweden}
\affiliation{ 
Department of Applied Physics, School of Engineering Sciences, KTH Royal Institute of Technology, Electrum 229, SE-16440 Kista, Sweden}%
\affiliation
{Department of Physics, University of Gothenburg, 412 96, Gothenburg, Sweden}

\date{\today}

\begin{abstract}
We study mutual synchronization in double nanoconstriction-based spin Hall nano-oscillators (SHNOs) under weak in-plane fields ($\mu_0H_\mathrm{IP}$~=~30--40~mT) and also investigate its angular dependence. We compare SHNOs with different nano-constriction spacings of 300 and 900~nm. In all devices, mutual synchronization occurs below a certain critical angle, which is higher for the 300~nm spacing than for the 900~nm spacing, reflecting the stronger coupling at shorter distances. Alongside the synchronization, we observe a strong second harmonic consistent with predictions that the synchronization may be mediated by the propagation of second harmonic spin waves. However, although Brillouin Light Scattering microscopy confirms the synchronization, it fails to detect any related increase of the second harmonic. Micromagnetic simulations instead explain the angular dependent synchronization as predominantly due to magneto-dipolar coupling between neighboring SHNOs.

\end{abstract}

\pacs{Valid PACS appear here}
\maketitle

\section{\label{sec:Intro}Introduction}

Nanoscale, spin current driven \cite{Berger1996,Slonczewski1996,Slonzewski1999jmmm}, auto-oscillators \cite{Tsoi1998,Myers1999,Kiselev2003,Rippard2004prl,Katine2008,Dumas2014,Silva2008,chen2016ieeerev} are of great interest both for ultra-wideband microwave signal generation \cite{bonetti2009, bonetti2010prl} and detection \cite{Tulapurkar2005nt}, and for neuromorphic computing \cite{Torrejon2017Nature,Romera2018nt,Grollier2016procieee}. A recent subset of these oscillators, so-called spin Hall nano-oscillators (SHNOs) \cite{Duan2014,Demidov2012,Pai2012,Demidov2014,Demidov2012b,Pai2012,Liu2013,Zholud2014,Collet2016,C6NR07903B,doi:10.1063/1.4907240,Ranjbar2014,Mazraati2016}, are driven by pure spin currents injected from an adjacent layer with strong spin-orbit coupling, such as Pt or W, and have demonstrated highly tunable microwave signals at room temperature with frequencies up to 28~GHz \cite{Zahedinejad2018apl}.

Nano-constriction based SHNOs \cite{Demidov2014} are particularly promising for applications as their fabrication is straightforward, their magnetodynamically active area is easily accessible for direct studies using optical access, and they exhibit very robust mutual synchronization \cite{Awad2016}, which improves their microwave signal properties by orders of magnitude. As mutual synchronization \cite{mancoff2005nt,kaka2005nt,sani2013ntc,Houshang2016,Urazhdin2016apl,Lebrun2017ntcomm} is also one of the most promising mechanisms for oscillator based neuromorphic computing \cite{Romera2018nt}, chains of mutually synchronized SHNOs \cite{Awad2016} will likely become the preferred implementation. 

However, mutually synchronized SHNOs have so far only been demonstrated in strong oblique magnetic fields of about 1 T with out-of-plane angles of around 80$^\circ$; for most applications this is impractical. Micromagnetic simulations also showed that mutual synchronization is promoted by the spatial expansion (delocalization) of the auto-oscillating mode, which happens when the frequency blueshifts with the bias current, which was initially believed to only occur in strong out-of-plane fields. As we have recently demonstrated theoretically \cite{dvornik2018anomalous} and experimentally \cite{mazraati2018} that auto-oscillations in nano-constriction SHNOs can also expand even in weak in-plane fields, it is interesting to explore the interaction between neighboring auto-oscillating regions also at these field conditions.

Here we experimentally demonstrate mutual synchronization of adjacent nano-constriction SHNOs in in-plane fields ($H_\mathrm{IP}$) as low as 0.03~T. We furthermore show that the coupling between the nano-constrictions, and hence their synchronization strength, can be tuned by the in-plane field angle such that strongest synchronization is achieved when the in-plane field is perpendicular to the line connecting the SHNOs. This is consistent with our micromagnetic simulations, which show that the auto-oscillating modes rotate to stay orthogonal to the field, thus offering different degrees of spatial overlap as a function of angle. As a consequence, mutual synchronization is only observed below a certain critical angle, which decreases with increasing SHNO separation.

When the SHNOs synchronize, we observe a concomitant substantial increase in the second harmonic consistent with predictions that the synchronization may be mediated by the propagation of second harmonic spin waves \cite{Kendziorczyk2016}. Although Brillouin Light Scattering microscopy confirms the synchronization, it however fails to detect any related increase of the second harmonic. Micromagnetic simulations instead explain the angular dependent synchronization as predominantly due to magneto-dipolar coupling between neighboring SHNOs.

\section{\label{sec:Exp}Experiment}
\subsection{\label{sec:DevFabrication}Device fabrication and measurement set-up}

\begin{figure*}
    \centering
\includegraphics[trim=0cm 0cm 1cm 0cm, clip=true,width=7in]{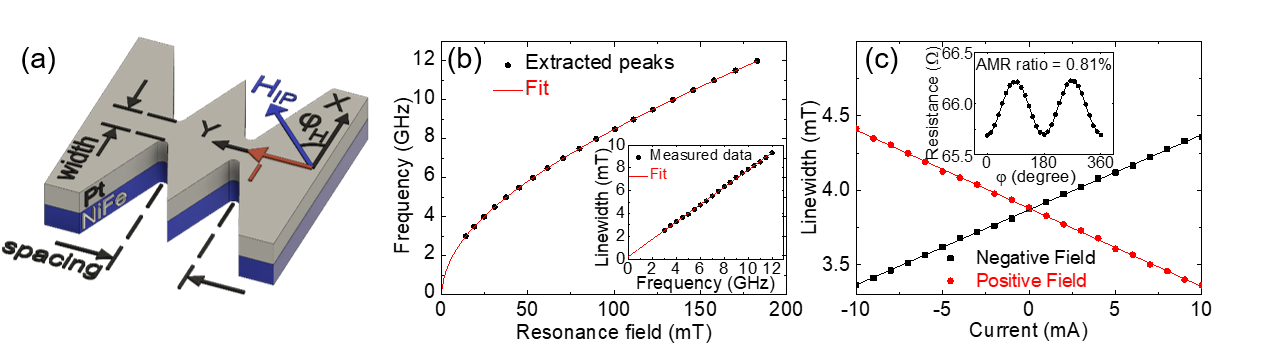}
\caption{\label{figFMR} (Color online) (a) Schematic of a double-NC SHNO and the direction of the in-plane field and current in our experiment (b) extracted ST-FMR peaks at $I_\mathrm{dc}$~=~0~mA for microwave frequencies from 3 to 12~GHz, and the Lorentzian fit to them. Inset: the extracted linewidth of the same spectra with the linear fit (c) extracted linewidth of ST-FMR peaks for $f$~=~5~GHz versus $I_\mathrm{dc}$ for both negative and positive field polarities along $\varphi$~=~30$^\circ$. Inset: the angular dependent magnetoresistance of the bar structure under $\mu_0H_\mathrm{IP}$~=~0.1~T}
\end{figure*}

SHNO stacks of NiFe(5nm)/Pt(5nm) were prepared on high-resistivity single crystal Si substrates using dc magnetron sputtering under a 3~mTorr Ar atmosphere in an ultra-high vacuum with $1.5\times 10^{-8}$~mTorr base pressure. Single and double nanoconstriction-based SHNOs with 150~nm width, separated by 300 to 900~nm, were defined using electron beam lithography, and the pattern was transferred to the stack using ion beam etching. The magnetodynamic properties of the bilayer were determined using spin-torque-induced ferromagnetic resonance (ST-FMR) measurements on 6~$\mu$m-wide bars fabricated next to the SHNO mesas. Finally, a conventional ground--signal--ground (GSG) waveguide and electrical contact pads for wide frequency range microwave measurement were fabricated by photolithography, followed by Cu/Au deposition and lift-off for both SHNOs and ST-FMR bars.

Figure~\ref{figFMR}(a) shows the schematic of a double-NC based SHNO and the configuration of the in-plane field ($H_\mathrm{IP}$) and the current: the field angle $\varphi$~=~$0^{\circ}$ is directed along the x-axis, while a positive current flows along the y-axis. ST-FMR measurements employing homodyne detection were carried out on bar-shaped devices by applying a 313 Hz pulse-modulated microwave signal alongside a dc current through a bias-tee. Using a lock-in amplifier, the modulated dc voltage was detected through the same bias-tee by sweeping the field from 200~mT to 0~mT, while keeping the frequency of the input microwave signal and the level of the input dc current fixed.

Microwave measurements were carried out in a custom-built setup \cite{Banuazizi2018rsi}. While a dc current was injected through the constriction area of the SHNO, the auto-oscillation microwave signal was captured using a spectrum analyzer, after amplification by 35 dB with a low-noise amplifier. 

The micro-focused Brillouin Light Scattering ($\mu$-BLS) measurements were performed using a 532 nm single-frequency (solid state diode-pumped) laser focused onto a diffraction-limited spot (360 nm) using a high numerical aperture objective (NA~=~0.75). The scattered light from the sample surface was then analyzed by a high-contrast six-pass Tandem Fabry–Perot TFP-1 interferometer (JRS Scientific Instruments). A nanometer-resolution stage allowed scanning of the sample around the constriction to make 2D spatial maps of the spin~waves. The sample was placed in the presence of a variable in-plane field magnet with the option of varying the field angle. The scattered light from the sample produced a BLS intensity proportional to the square of the amplitude of the magnetization dynamics at the corresponding frequency. The $\mu$-BLS set-up is equipped with a spectrum analyzer connected to the sample via a bias-tee to measure the electrical and the optical signals simultaneously.
All the measurements were performed at room temperature.

\subsection{\label{sec:ElectricalResults}Electrical microwave spectroscopy}

Figure~\ref{figFMR}(b) shows the ST-FMR frequency versus the resonance field of the 6~$\mu$m-wide bar with no bias current. The resonance field values were extracted by fitting each peak to the sum of a symmetric and an asymmetric Lorentzian function.\cite{Liu2011a} The frequency vs. field dependence follows the Kittel formula\cite{Kittel1948} (solid red line) with an effective magnetization of $\mu_{0}M_\mathrm{eff}$~=~0.80~T and gyromagnetic ratio of $\gamma/2\pi$~=~28.3~GHz/T. The inset shows the frequency dependence of the extracted linewidth for the same data set. The linear fit to the linewidth (solid red line)\cite{Kittel1948} gives the Gilbert damping constant value of $\alpha$~=~21.5~$\times$~10$^{-3}$. 

Figure~\ref{figFMR}(c) shows the bias current, $I_\mathrm{dc}$, induced changes in the linewidth of the ST-FMR spectra arising from the spin-orbit torque. The data was measured at the microwave frequency of 5~GHz for positive and negative in-plane magnetic field directions. Both datasets show linear trends, as highlighted by the solid lines. Passing a positive dc current decreases (increases) the linewidth for positive (negative) polarity of the applied field. Using the slope of the linewidth versus the current, the spin Hall efficiency---defined as the ratio of the absorbed spin to charge current densities, $\xi_\mathrm{SH}=J_\mathrm{s}/J_\mathrm{c}$~\cite{Demasius2016,Ando2008c,Liu2011a}, is calculated as 0.085. Finally, an AMR value of 0.81 $\%$ is derived from the changes in the magnetoresistance of the bar versus $\varphi$ under $\mu_0H_\mathrm{IP}$~=~0.1~T, as shown in the inset of Figure~\ref{figFMR}(c).

\begin{figure}[t]
    \centering
\includegraphics[trim=0.7cm 0cm 0.9cm 0cm, clip=true,width=3.4in]{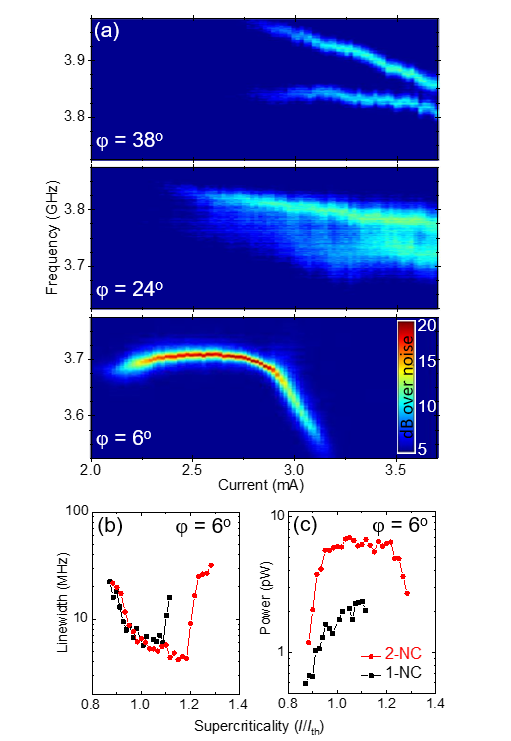}
\caption{\label{figCurrent} (Color online) (a) Current-dependent auto-oscillation PSDs of a 150~nm-wide double-NC SHNO with 300~nm spacing at $\mu_0H_\mathrm{IP}$~=~30~mT directed along $\varphi$~=~38$^{\circ}$, 24$^{\circ}$, and 6$^{\circ}$. (b) The extracted linewidth and (c) the integrated power for single- and double-NC SHNOs at $\mu_0H_\mathrm{IP}$~=~30~mT directed along $\varphi$~=~6$^{\circ}$.}
\end{figure}

We then carried out microwave characterization of the current-driven 150~nm-wide SHNOs pair separated by 300~nm. Figure~\ref{figCurrent}(a) shows the auto-oscillation power spectral density (PSD) vs.~ current for the field applied along $\varphi$~=~38$^{\circ}$, $24^{\circ}$, and $6^{\circ}$. At the highest angle, we observe two spin wave modes separated by roughly 125~MHz at their auto-oscillation onsets. This is the unsynchronized state with difference in free running frequencies of SHNOs stemming from the constrictions' fabrication-related shape and width variations. The modes also show weak, but different values of nonlinearly frequency red shifts indicative of the linear-like type of auto-oscillations\cite{mazraati2018}. Distinct frequency vs. current slopes allow the modes to approach each other as close as 50~MHz at the highest bias current of 3.7~mA. However, the SHNOs remain unsynchronized suggesting a locking bandwidth lower than 50~MHz for the applied field angle of $\varphi$~=~38$^{\circ}$.

A similar behavior is observed for the field applied at $\varphi$~=~24$^{\circ}$. Compared to $\varphi$~=~38$^{\circ}$ the frequencies difference at the onset of the auto-oscillations reduces to approximately 60~MHz and remains constant with the bias current. This is in contrast to the case of out of plane applied fields where robust mutual synchronization was already observed at such in-plane field angles. This could be attributed to the non-monotonic frequency vs. current behaviour that enables the frequency crossing and frequency blue shifting that enhances coupling strength\cite{Awad2016}.

By decreasing the in-plane field angle further to $\varphi$~=~$6^{\circ}$, we observe only one auto-oscillating mode right from the threshold current, $I_\mathrm{th}$. Now the frequency vs. current behaviour is non-monotonic, changing from blue- to red-shifting at around 2.6~mA. Then at roughly 2.9~mA, the red-shift increases substantially, the amplitude of the signal reduces and the linewidth increases, suggesting condensation of the linear-like modes to spin wave bullets\cite{mazraati2018}. Despite the change of dynamics, we still measure a single auto-oscillating mode. 

Figures~\ref{figCurrent}(b) and (c) show the auto-oscillation linewidth and integrated power versus supercriticallity, $I/I_\mathrm{th}$, of a single (black) and double (red) nano-constriction SHNO. The threshold current for each SHNO is extracted from linear fits of the inverse integrated power vs.~current\cite{Tiberkevich2007c,Slavin2009a}. We compare the signals at the same values of the supercriticality to ensure comparable (a) intrinsic powers of the oscillators and (b) nonlinear linewidth amplification effects\cite{Kim2008prl}. 

From now on we will focus on the linear-like regime that shows the highest output power and signal coherence. First of all, we note that the microwave signal from the double nano-constriction shows lower linewidth above the auto-oscillation threshold. It keeps decreasing vs.~supercriticallity down to the theoretical limit of the synchronized state of half generational linewidth of a single oscillator. Secondly, we observe that the output power of the pair, $P_{2}$ is fairly constant above threshold current at around 6.3~pW. The power of the single SHNO, $P_{1}$ slowly increases with supercticallity approaching a value of 2.4~pW at the highest measured current. At this point, a pair of SHNOs shows an output power enhancement by a factor of $P_{2}/P_{1}=2.63$.

At first look, the observed power scaling significantly exceeds a theoretical limit of $P_{2}/P_{1}=2$. However, it should be noted, that the output power reported here is measured on the $Z_{0}$~=~50~$\Omega$ load, i.e., the so-called delivered power. All our devices measure more than 150~$\Omega$ resistance and so, there is significant power loss due to the resistance mismatch. The scaling of the delivered power could be estimated using the model developed in Ref. \cite{georges2008impact}. For a pair of in-phase synchronized oscillators, it reads $P_{2}/P_{1} = 4(Z_{0}+Z_{1})^{2}/(Z_{0}+Z_{2})^{2}=2.7$, where $Z_{1}$~=~180~$\Omega$ and $Z_{2}$~=~230~$\Omega$ are resistances of a single and a pair of SHNOs, respectively. Here we assumed that the AMR ratio and intrinsic power of SHNO does not change with the number of oscillators. 

Since the observed delivered power scaling agrees well with the theoretical value, we conclude that for the in-plane field applied at $\varphi$~=~$6^{\circ}$ we observe synchronization of SHNOs pair with vanishing phase lag. In contrast, it approached tens of degrees for the out-of-plane fields \cite{Awad2016}. So, rf-applications-wise the phase-locking in weak in-plane fields is preferable as vanishing phase lag would enable linear scaling of the generated power with number of oscillators. 

We note that our measurements also suggest synchronization of SHNOs in the strongly nonlinear regime (i.e, beyond $I_\mathrm{dc}$~=~2.9~mA). This might be the very first experimental demonstration of phase locking of spin wave bullet solitons. However, a much worse quality of fittings in this regime does not allow us to estimate the linewidth and power scaling and give a definite answer at this point.

\begin{figure}[t]
    \centering
\includegraphics[trim=0cm 0cm 0cm 0cm, clip=true,width=3.4in]{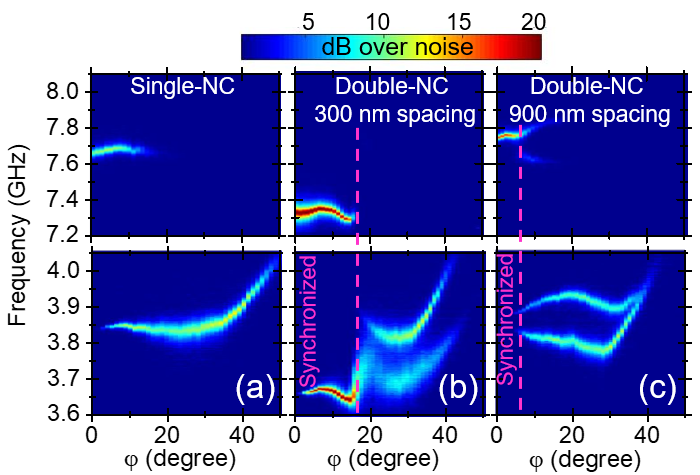}
\caption{\label{figTheta} (Color online) Fundamental and second harmonics of the auto-oscillation PSD versus field angle at $\mu_0H_\mathrm{IP}$~=~30~mT and $I_\mathrm{dc}$~=~2.8~mA for (a) single-NC, (b) double-NC with 300~nm spacing, and (c) double-NC with 900~nm spacing}
\end{figure}

As synchronization favours lower in-plane applied field angles, we look at its angular dependence more closely. Figure \ref{figTheta} shows spectra of (a) a single, and (b) a pair of 300~nm separated SHNOs vs.~field angle ($\mu_0 H$~=~30~mT) at a bias current of $I_\mathrm{dc}$~=~2.8~mA. At lower angles, the fundamental mode (see bottom row) could not be detected electrically, since the first derivative of the AMR curve is vanishing (Fig.~\ref{figFMR}(c), inset). At the same time, a strong second harmonic appears at lower angles (see top row) as the second derivative of the AMR curve is large at this region.\cite{Muduli2011jap} Since the amplitude of the second harmonic is comparable to the fundamental mode, it could be used for relatively high-frequency rf signal generation even in weak applied fields. At the same time, it lies well above FMR and, thus, might directly excite propagating spin waves\cite{Kendziorczyk2016, demidov2016excitation}, \emph{e.g.}~act as a nano-scale spin wave source for the applications in magnonics.

The single SHNO produces predominantly the second harmonic for $\varphi \lesssim 5^{\circ}$, Fig. \ref{figTheta}(a), then in a range of angles $5^{\circ} \lesssim \varphi \lesssim 15^{\circ}$ a signal at the fundamental frequency appears with comparable amplitude and persists up until $\varphi$~=~$50^{\circ}$, while the second harmonic vanishes for $\varphi \gtrsim 15^{\circ}$. For low in-plane field angles, the fundamental signal shows a weak nonmonotonic frequency vs.~angle behavior and then monotonically increases from 3.85~GHz at $\varphi \approx 25^{\circ}$ to 4.05~GHz at $\varphi \approx 50^{\circ}$, suggesting SHNO operation in a linear-like mode\cite{mazraati2018}. 

For the SHNO pair with 300~nm spacing (Fig.~\ref{figTheta}(b)), the constrictions oscillate independently with different frequencies at higher angles, similar to what we observed in current sweeps shown in Fig.~\ref{figCurrent}(a). Moving towards smaller angles, abrupt synchronization happens below a critical angle of $\varphi_\mathrm{crit}\simeq20^{\circ}$, as indicated by the significantly increased amplitudes of the fundamental and second harmonics signals. The former disappears in the vicinity of $\varphi = 0^{\circ}$, and a pair of SHNO only delivers power at the double frequency. 

At the synchronization point, the locking bandwidth must exceed the difference in constrictions' free running frequencies of 120~MHz. Since it is fairly constant above the critical angle, we conclude that locking bandwidth is applied-field-angle-dependent. In general, the spatial auto-oscillation profile is elongated perpendicular to the applied filed direction\cite{Demidov2014, mazraati2018}, \emph{e.g.}~along the SHNO stacking direction for the field applied perpendicular to the constriction. So the locking bandwidth decreases with increasing angle due to the reduced spatial overlap of the auto-oscillating regions\cite{Kendziorczyk2016}. Mutual synchronization was also achieved for a triple-constriction SHNO with 300~nm spacing. Since no fundamental changes to the synchronization behaviour were observed in that case, we decided to not discuss it in details here.

Increasing the constriction spacing from 300~nm to 900~nm (Fig.~\ref{figTheta}(c)) pushes the critical angle down to $\varphi_\mathrm{crit} \simeq5^{\circ}$. Consequently, only a strong second harmonic of the synchronized state is observed. It should be noted that in this case, the locking bandwidth is around 60~MHz, albeit of the expected strong spatial overlap of the modes. Also at the field angles of around $\varphi \approx 40^{\circ}$, the signals of free running SHNOs cross, yet no signs of synchronization are observed. So we conclude that increasing the constriction spacing from 300~nm to 900~nm significantly reduces their mutual coupling and, therefore, their locking bandwidth. This is expected irrespective of the interaction mechanism.

\begin{figure}[t]
    \centering
     \includegraphics[trim=0cm 0cm 0cm 0cm, clip=true,width=3.4in]{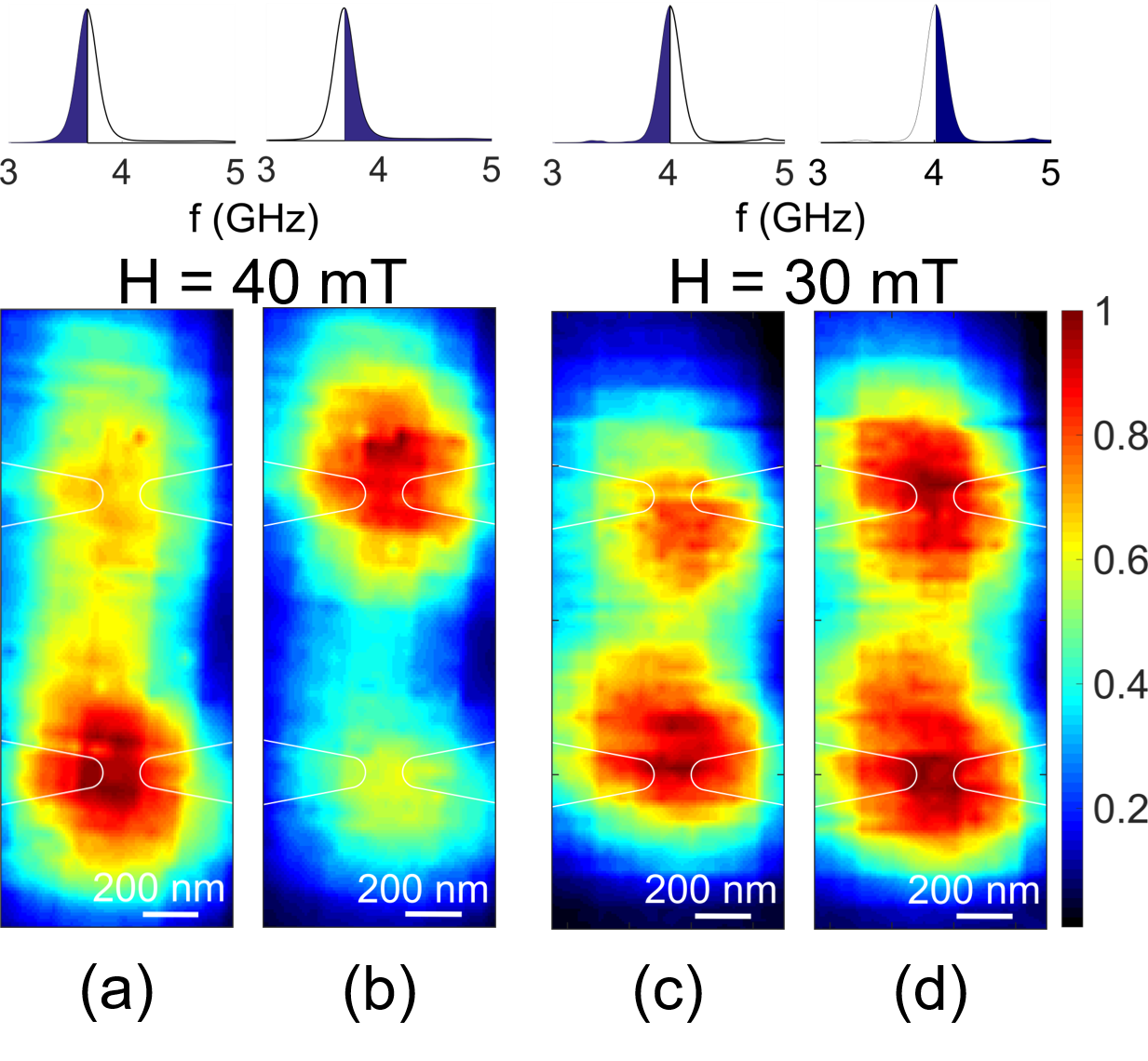}
\caption{\label{figBLS1} (Color online) (a--b) Spatial profile of the oscillation modes for unsynchronized oscillators at $I_\mathrm{dc}$~=~3~mA and $\mu_0H_\mathrm{IP}$~=~40~mT where the oscillators are seen to be operating at separate frequencies. (c--d) the spatial profile of the oscillation modes for synchronized oscillators at $\mu_0H_\mathrm{IP}$~=~40~mT, showing two constrictions operating at the same frequency. The frequency range of each of the spatial profiles are given by the shaded region in the spectra.}
\end{figure}

\subsection{\label{sec:BLSresults}Brillouin Light Scattering microscopy}

In contrast to electrical measurements that rely on the AMR effect, the Brillouin light scattering (BLS) sensitivity is independent of the in-plane field angle. So we carried out optical measurements to directly inspect the spatial properties of the auto-oscillating spin wave modes. A raster scan across the sample using BLS gave a clear picture of the size and shape of the modes. 

Figure~\ref{figBLS1} shows the spatial profiles of the modes in a device with an interconstriction distance of 900~nm at $I_\mathrm{dc}$~=~3~mA for fields applied along an in-plane angle of $\varphi$~=~3$^\circ$. The larger separation was chosen to be longer than the spatial resolution of the BLS ($\sim$~360~nm), in order to resolve each constriction separately. Due to the limited spectral resolution of the measurement (of around 100~MHz), we observe a single frequency peak for both synchronized and unsynchronized states (Fig.~\ref{figBLS1}, top row). However, to recognize the synchronization, we made 2D spatial maps of the auto-oscillations by integrating either the upper half or the lower half of the resonant peak; the corresponding frequency ranges are shown by the blue shaded area in the spectra. In the unsynchronized state at $\mu_0H_\mathrm{IP}$~=~40~mT (Fig.~\ref{figBLS1}~(a--b)), each individual oscillator operates at slightly different frequencies. So the spatial maps, Fig. \ref{figBLS1}(a) and (b), show auto-oscillations power mostly concentrated to either of the two constrictions. For the synchronized state, at the lower field of 30~mT (Fig.~\ref{figBLS1}~(c--d)), the modes are inseparable in frequency so, that the spatial maps show even distribution of the auto-oscillations power across the pair. We also note that, compared to the unsynchronized state, the modes appear enlarged and shifted towards each other. Some accommodation of the spatial mode profile is likely required to make the frequencies overlap.

\begin{figure}[t]
    \centering
     \includegraphics[width=3.4in]{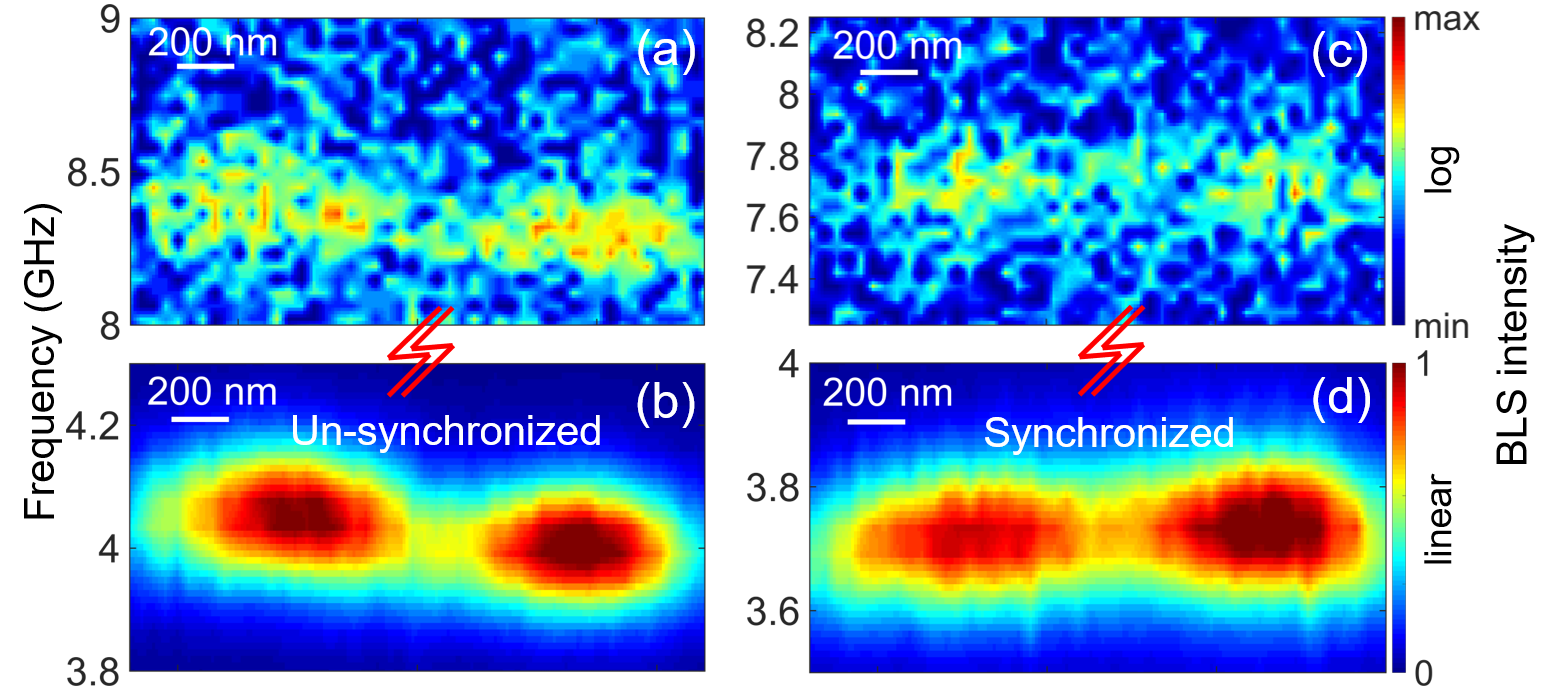} 
\caption{\label{figBLS2} (Color online) Line scans along the constrictions showing both first (b \& d) and second harmonic modes (a \& c) for the synchronized and unsynchronized states.}
\end{figure}

Extracting the spectrum along the line connecting the two constrictions, Fig.~\ref{figBLS2}, shows that (d) the synchronized oscillators oscillate at the same frequency, (b) while the unsynchronized devices are separated by about 50~MHz. We also detect a weak magnetodynamical signal at the second harmonic, Fig.~\ref{figBLS2}(a) and (c), which lies well above FMR, $\approx 5$~GHz and so, must correspond to propagating spin waves. According to the micromagnetic simulations of comparable devices, the wavelength of such magnons should be less than 200~nm\cite{Kendziorczyk2016}. So the relatively low amount of BLS counts detected could be due to the reduced sensitivity of our BLS setup to spin wave wavelengths well below 360~nm.

\section{\label{sec:Simulation}Micromagnetic simulations}

To gain a deeper insight into the auto-oscillations, their dynamics, and their synchronization, we carried out micromagnetic simulations using the GPU-accelerated program \textsc{mumax3}.\cite{Vansteenkiste2014} We simulated a pair of 140~nm and 160~nm wide constrictions separated by 300~nm with a mesa area of 2000$\times$4000~nm$^2$ and a thickness of 5~nm. We only modeled the ferromagnetic layer, while the effect of the heavy metal was included by considering a spin current perpendicular to the magnetic layer due to the spin-Hall conversion of charge current in the Pt layer. Its distribution and total Oersted field at the NiFe site were calculated using COMSOL Multiphysics$\textsuperscript{\textregistered}$ software\cite{Dvornik2018prappl} for NiFe(5~nm)/Pt(5~nm) bilayer, using resistivities of 90~$\mu\Omega$.cm and 44~$\mu\Omega$.cm for NiFe and Pt, respectively. 

The saturation magnetization, spin-Hall efficiency, gyromagnetic ratio, and damping parameter were obtained from the ST-FMR measurement. We also implemented an absorbing periodic boundary condition in order to avoid auto-oscillations self-locking to the spin waves excited at double the oscillations frequencies and back-reflected from the mesa boundaries. The exchange stiffness was set at a typical value for NiFe layer $A_\mathrm{ex}$~=~10~pJ/m.\cite{coey_magnetism_2010} In the simulations, we first allowed the magnetization to relax to its minimum energy state, then applied the current and let the system evolve for up to 300~ns. We then carried out spectral analysis of the $y$~component of the magnetization over the period of [100-300]~ns to obtain the auto-oscillation spectra and mode profiles, excluding the transient effects. All simulations were done at zero temperature.

\begin{figure}[t]
    \centering
\includegraphics[width=3.4in]{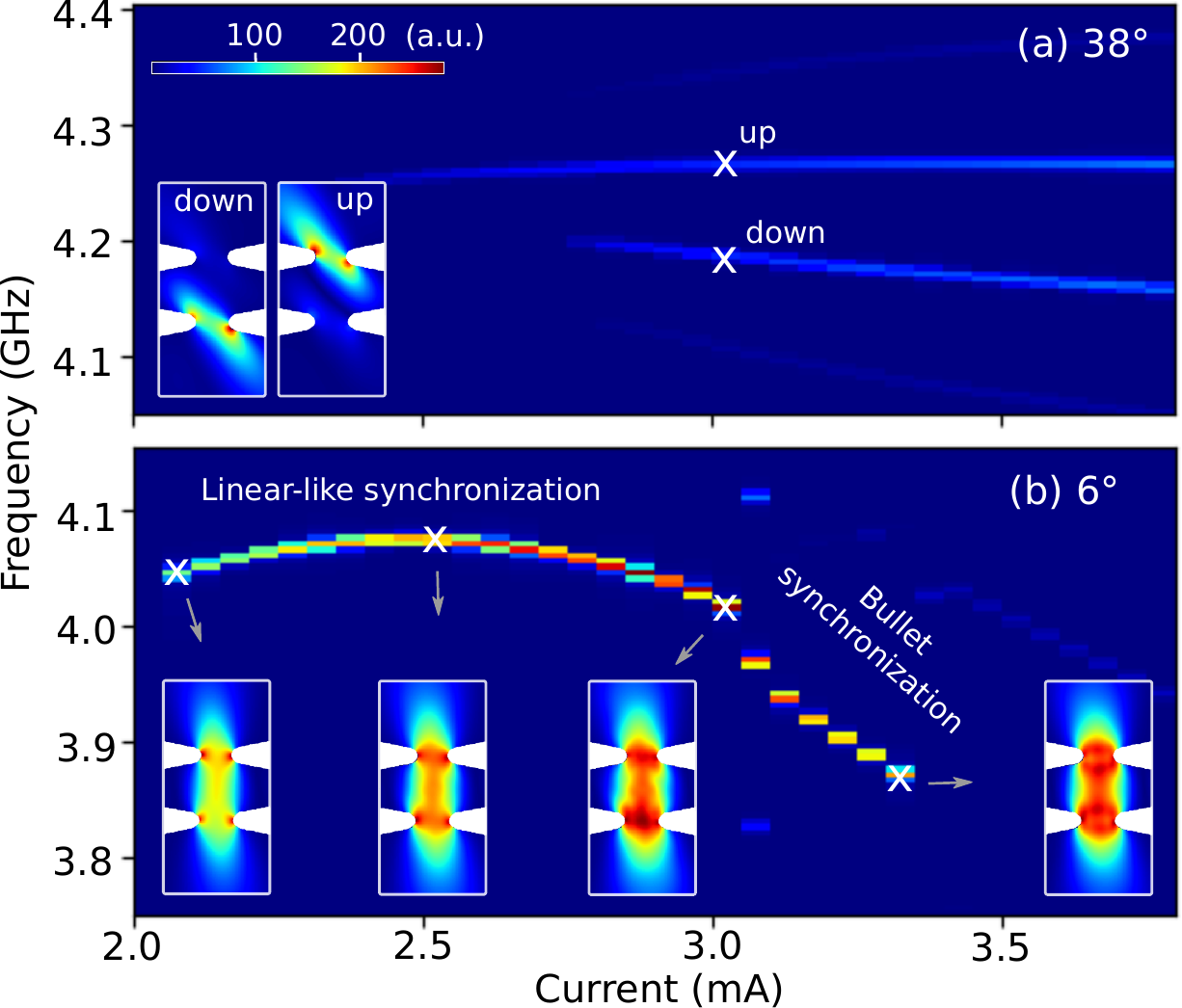}
\caption{\label{Simulation_curren_scan} (Color online)(Simulation) The FFT amplitude of total magnetization versus DC current at $\mu_0H_\mathrm{IP}$~=~28~mT for (a) $\varphi$~=~$38^{\circ}$, and (b) $\varphi$~=$~6^{\circ}$. The insets show the spatial distribution of the FFT of the local magnetization at the excited frequency (known as the mode profile) corresponding to the currents employed. In each mode profile, the red (blue) color corresponds to the maximum (minimum) of the FFT amplitude at the specific current, angle, and frequency. }
\end{figure}

Figures~\ref{Simulation_curren_scan}~(a) and (b) compare the current scan simulations for $\varphi$~=~$6^{\circ}$ and $\varphi$~=~$38^{\circ}$, respectively, corresponding to the experimental results shown in Fig.~\ref{figCurrent}; the simulations agree well with the measurements. First of all, a 20~nm variation in SHNOs widths is sufficient to reproduce the experimentally observed frequency difference of unsynchronized SHNOs at $\varphi$~=~$38^{\circ}$. The extracted mode profiles (see insets) show that devices oscillate independently in a linear-like regime\cite{mazraati2018}, and the higher generation frequency corresponds to the smaller constriction. Secondly, at $\varphi$~=~$6^{\circ}$ oscillations show a nonmonotonic frequency vs.~current behavior with complete synchronization within the inspected range of the electrical current. At the onset of the auto-oscillations, $I_\mathrm{dc}$~=~2.05~mA, the dynamics is linear-like followed by a slight frequency blue shifting and substantial mode expansion as the current increases to $I_\mathrm{dc}$~=~2.5~mA. The nonlinearity then turns negative, resulting in a transition to the bullet mode at around 3~mA with a stronger negative current dependence of the frequency. Despite the mode transition, the oscillators remain synchronized, confirming the experimentally suggested phase-locking of spin wave bullet solitons. Contrary to our previous report on the bullets' volume reduction vs.~current in a single constriction SHNO\cite{mazraati2018}, the collective synchronized bullet mode appears to expand as the current increases from 3~mA to 3.4~mA. Synchronization of magnetodynamic solitons may hence very well have some unexpected consequences, although this is beyond the scope of the present manuscript.

\begin{figure}[t]
    \centering
\includegraphics[width=3.4in]{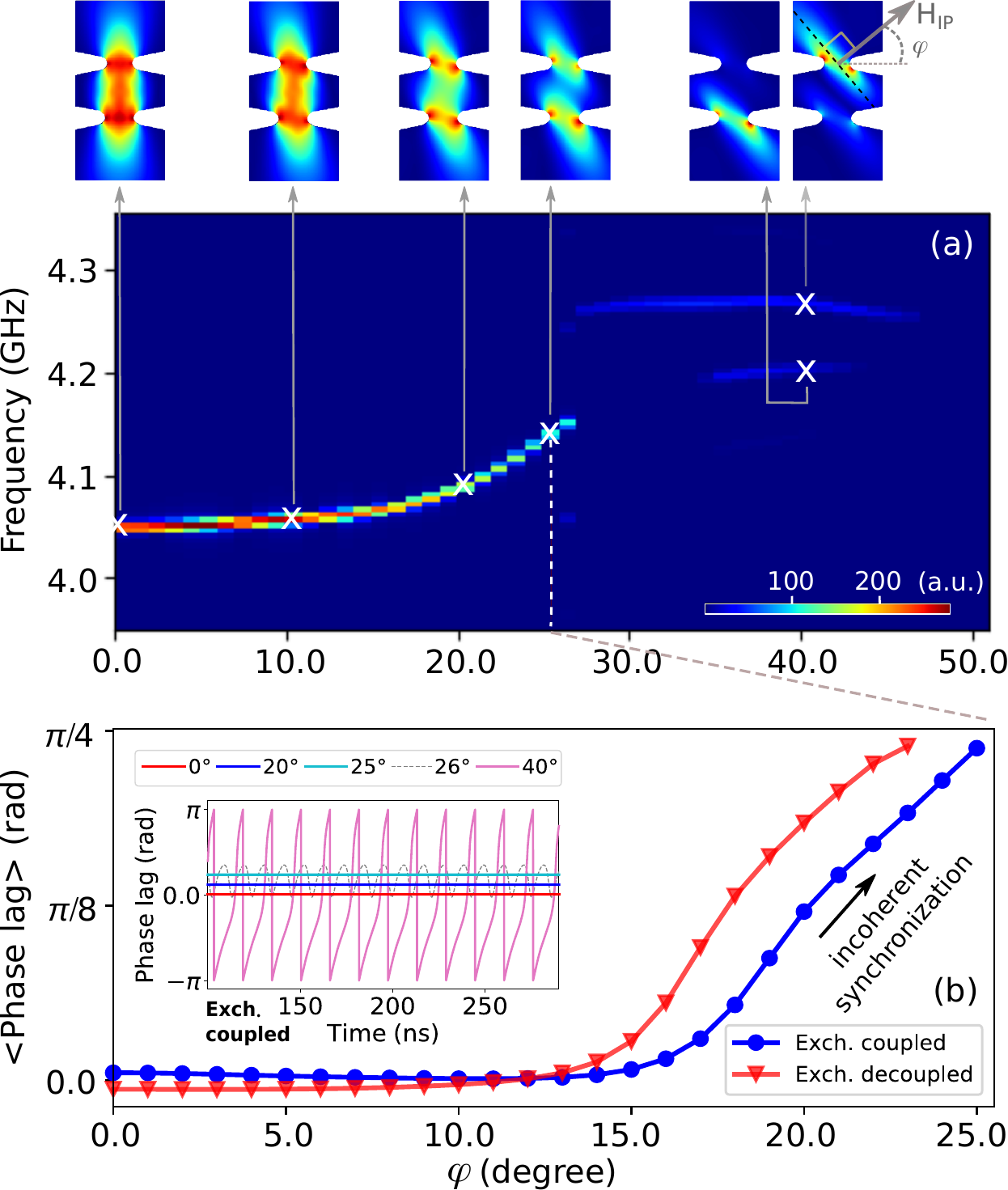}
\caption{\label{Simulation_angle_scan} (Color online)(Simulation) (a) FFT amplitude of the total magnetization versus the in-plane angle of the external magnetic field with a magnitude of $\mu_0H_\mathrm{IP}$~=~28~mT at $I_\mathrm{dc}$~=~2.8~mA. The mode profiles corresponding to the given in-plane angles are plotted above the FFT amplitude spectral density (in each mode profile, the red (blue) color corresponds to the maximum (minimum) FFT amplitude at the specific given current, angle, and frequency). (b) The time-averaged phase lag between the two nano-oscillators in the synchronized regime for normal (blue) and exchange decoupled (red) structures. The inset shows the time evolution of the phase lag for different in-plane angles. Calculating the variation of the sine of phase lag, we find it to be near zero at synchronized angles ($0^{\circ}$, $20^{\circ}$, and $25^{\circ}$) and around 0.5 for the unsynchronized angle ($40^{\circ}$).\cite{Vassilieva2011ieeeTNN} } 
\end{figure}

Figure~\ref{Simulation_angle_scan}~(a) shows the in-plane angular scan simulated for $I_\mathrm{dc}$~=~2.8~mA and $\mu_0H_\mathrm{IP}$~=~28~mT, which correspond to the linear-like auto-oscillations. Similar to the experimental observations (Fig.~\ref{figTheta}~(b)), as the in-plane angle decreases, the auto-oscillations enter the synchronized regime below a critical in-plane angle of around $\varphi_\mathrm{crit} \simeq25^{\circ}$. The extracted mode profiles shown in Fig.~\ref{Simulation_angle_scan} confirm the rotation of the auto-oscillations with the applied field in-plane angle. At low angles, $\varphi$~=~$0^{\circ}$ and $10^{\circ}$, the auto-oscillations not only occupy most of the constrictions space, but also extend well into the bridge connecting the SHNOs and, overall, appear as a single mode. For the moderate angles, $\varphi$~=~$20^{\circ}$ and $25^{\circ}$, the auto-oscillations are significantly more localized to the edges of the constriction, and their amplitude in the bridge is much reduced due to the rotation. Finally, above the critical angle, $\varphi$~=~$25^{\circ}$, there is vanishing overlap between the modes.

The coherence of the synchronized state is determined by the phase lag between the oscillators. We, therefore, employ a Hilbert transform of the time-domain signals from each oscillator of the pair and estimate their phase difference as functions of time and in-plane field angle. An example of such analysis is shown in the inset of Fig.~\ref{Simulation_angle_scan}(b). For the applied field angles of $\varphi$~=~$0^{\circ}$, $20^{\circ}$, and $25^{\circ}$, the phase lag is constant with time, as expected for synchronized oscillators without noise sources that might induce phase slips. Above the critical angle, there are periodic phase slips since oscillators are not synchronized. 

The blue curve in Fig.~\ref{Simulation_angle_scan}(b) shows the mean phase lag between the synchronized auto-oscillating constrictions vs.~applied field angle. While the phase lag shows a non-monotonic behaviour with a minimum at around $\varphi$~=~$12^{\circ}$, it is virtually zero up to $\varphi$~=~$15^{\circ}$. Assuming that the difference in free running frequencies of the oscillators weakly depends on the applied field angle, we conclude that the coupling between the auto-oscillations is rather strong in this range of angles. Above $\varphi$~=~$15^{\circ}$, the phase lag rapidly increases, consistent with a reduced coupling originating from the reduced spatial overlap of the modes. 

In principle, the magnetic coupling of the constriction-based SHNOs could arise from three distinct mechanisms: magneto-dipolar, direct exchange and propagating spin waves; coupling through the electrical current should be negligible due to the very low magnetoresistance. The last two magnetic types require exchange coupling between the devices, which can be suppressed by making a trench in the center of the bridge connecting the constrictions, similar to what was done by Pufall \textit{et al.}~to elucidate the interaction mechanism in nanocontact spin torque nano-oscillators\cite{Pufall2006prl}. While this method is not easily implemented in our experiments, since the bridge provides the electrical connection between the devices, simulations can directly determine the impact of exchange coupling. We, therefore, removed the exchange coupling between the top and bottom constrictions and repeated the angular-resolved phase lag analysis; the corresponding data is shown by the red curve in Fig.~\ref{Simulation_angle_scan}(b). The effect of exchange decoupling is negligible up to $\varphi$~=~$12^{\circ}$. After that, the phase lag rapidly increases leading to a $2^{\circ}$ (8\%) reduction in the critical synchronization angle, as compared to the unconstrained simulations. So the direct exchange coupling, which is completely eliminated in our simulations, does not play a predominant role in the observed synchronization. 

Kendziorczyk and Kuhn showed that spin-wave-mediated coupling dominates at larger distances (800~nm for devices similar to ours). In such a regime, the phase-locking bandwidth shows oscillatory behaviour vs.~relative shift between the constrictions. In particular, the phase-locking bandwidth should be enhanced if spin waves arrive at opposite devices with zero phase shift. The apparent lack of oscillatory behaviour vs.~applied field angle in our simulations indicates vanishing contribution of the spin waves to the synchronization. Furthermore, the exchange decoupling that we implemented in simulations should attenuate and shift phase of the spin waves traveling through the center of the bridge. Again, our simulations show no qualitative changes in this case. So, we conclude that the magneto-dipolar coupling is the dominant coupling mechanism for the in-plane magnetized constriction-based SHNOs.

\section{\label{sec:Conclusion}Conclusion}

We have presented a detailed study of mutual synchronization of nanoconstriction-based SHNOs in weak in-plane fields of 30--40~mT with particular emphasis on the dependence of synchronization on the field angle. We studied double-nanoconstrictions with spacings of 300 and 900~nm. Synchronization was achieved at angles below a critical angle, $\varphi_\mathrm{crit}$, which is higher for the shorter spacing due to the stronger coupling between the nanoconstrictions. The spatial profiles from the micromagnetic simulations show that the synchronization at lower angles is due to the expansion of the field-localized spin wave modes in a direction perpendicular to the external field. When the SHNOs synchronize, we observe a substantial increase in the second harmonic. While this observation is in principle consistent with predictions that the synchronization may be mediated by the propagation of second harmonic spin waves, both Brillouin Light Scattering microscopy and micromagnetic simulations rule out this mechanism. Instead, the angular dependent synchronization is predominantly governed by magneto-dipolar coupling between neighboring SHNOs.

\section*{Acknowledgments}

This work was supported by the Swedish Foundation for Strategic Research (SSF), the Swedish Research Council (VR), and the Knut and Alice Wallenberg foundation (KAW). This work was also supported by the European Research Council (ERC) under the European Community's Seventh Framework Programme (FP/2007-2013)/ERC Grant 307144 ``MUSTANG''.

\end{document}